\documentclass[12pt]{article}
\usepackage{epsfig}

\def\be{\begin{equation}}
\def\ee{\end{equation}}
\def\ben{\begin{displaymath}}
\def\een{\end{displaymath}}
\def\ba{\begin{array}{c}}
\def\bal{\begin{array}{l}}
\def\ea{\end{array}}
\def\p{\partial}
\begin{document}

\titlepage

 \begin{center}
{\tiny .}

\vspace{.35cm}

{\Large \bf

Matrix Hamiltonians with an algebraic guarantee of unbroken ${\cal
PT}-$symmetry

   }\end{center}

\vspace{10mm}

 \begin{center}

 {\bf Miloslav Znojil}

 \vspace{3mm}
Nuclear Physics Institute ASCR,

 250 68 \v{R}e\v{z}, Czech Republic

{e-mail: znojil@ujf.cas.cz}

\vspace{3mm}

\vspace{5mm}
%

\end{center}


\vspace{5mm}

\section*{Abstract}

Although the quantum bound-state energies may be generated by the
so called ${\cal PT}-$symmetric Hamiltonians $H = {\cal
P}H^\dagger{\cal P}\neq H^\dagger$ where ${\cal P}$ is, typically,
parity, the spectrum only remains real and observable (i.e., in
the language of physics, the ${\cal PT}-$symmetry remains
unbroken) inside a domain ${\cal D}$ of couplings. We show that
the boundary $\p {\cal D}$ (i.e., certain stability and
observability horizon formed by the Kato's exceptional points)
remains algebraic (i.e., we determine it by closed formulae) for a
certain toy-model family of  $N-$dimensional
anharmonic-oscillator-related matrix Hamiltonians  $H^{(N)}$ with
$N=2,3,\ldots,11$.

\newpage

\section{Introduction}

According to the abstract principles of Quantum Mechanics, the
observable quantities (say, the spectra of energies
$E_0<E_1<\ldots$ of bound states) should be constructed as
eigenvalues of a certain self-adjoint operator $H=H^\dagger$
acting in some physical Hilbert space of states ${\cal H}$.
Fortunately, the full and impressive generality of this
formulation of the theory is rarely needed in its concrete
applications. Most often, the Hilbert space is being chosen in its
most common representation $I\!\!L_2(I\!\!R)$ with elements
$|\psi\rangle$ representing the square-integrable complex
functions of a single variable $x$ interpreted as a coordinate of
a (quasi)particle.

The most common version of the Hamiltonian $H$ composed of its
kinetic and potential-energy parts leads to the constructions of
the energies via a suitable phenomenological potential $V(x)$
entering the ordinary differential Schr\"{o}dinger equation
 \be
 -\frac{\hbar^2}{2m}\,\frac{d^2}{dx^2}\,\psi^{}(x)
 + V^{}(x)\,\psi^{}(x) = E\,\psi^{}(x)\,.
 \label{SE}
 \ee
In 1998, Bender and Boettcher \cite{BB} demonstrated that the
spectra $\{E_n\}$ can remain real, amazingly, even if the
potentials themselves become complex. This attracted the attention
of physicists of different professional orientations ranging from
supersymmetry \cite{Cannata} and field theory \cite{BM} to
cosmology \cite{Alifirst} and even
magnetohydrodynamics~\cite{Uwe}.

Although the Bender's and Boettcher's observation has been
supported by many subsequent studies and concrete
examples~\cite{DDT,BBb,solvable} it still looked like a paradox as
it immediately implied that $H \neq H^\dagger$ in
$I\!\!L_2(I\!\!R)$. Fortunately, the resolution of the paradox
proved rather easy \cite{BBJ}. It was sufficient to imagine that
the Hamiltonian $H$ can remain tractable as self-adjoint in {\em
another} Hilbert space of states ${\cal H}\neq I\!\!L_2(I\!\!R)$.
A closely related {\em historical} paradox is that the usefulness
of transfer of $H\neq H^\dagger$ to another Hilbert space ${\cal
H}$ (where it becomes self-adjoint) has already been known and
even applied successfully in nuclear physics in 1992 \cite{Geyer}.
Still, it took time to clarify this parallelism \cite{proceedings}
(cf. also the comprehensive review paper \cite{Carl} for further
details).

On the level of the elementary ordinary differential example
(\ref{SE}), one of the most important formal observations has been
made in ref.~\cite{DDT}. It's authors Dorey, Dunning and Tateo
noticed and described a {\em deep nontriviality} of the problem of
the reality of the spectrum $\{E_n\}$. After they added some other
free parameters (the physical meaning of which is not too relevant
for our forthcoming argumentation) and after they replaced the
Bender's and Boettcher's $H^{(BB)}= p^2+x^2\,({\rm i}x)^\delta$
with $\delta\geq 0$ by a two-parametric family of their
generalized Hamiltonians $H^{(DDT)}(\alpha,\ell)$, they revealed
that the reality of the spectrum $\{E_n(\alpha,\ell)\}$ only takes
place {\em inside a certain domain} of parameters [let's denote it
as ${\cal D}(\alpha,\ell)$] with a highly nontrivial, spiked shape
of its boundary $\partial {\cal D}(\alpha,\ell)$.

We found the latter observation extremely challenging, important,
interesting and inspiring. The very existence of a finite boundary
of the domain ${\cal D}(\alpha,\ell) \neq I\!\!R^2$ represents one
of the main differences of the model from the differential
Schr\"{o}dinger eqs.~(\ref{SE}) which are self-adjoint in the
standard Hilbert space $I\!\!L_2(I\!\!R)$ and which require that
the potential $V(x)$ remains real. In the new framework, the
change of space $I\!\!L_2(I\!\!R) \to {\cal H} $ implies that only
the new potentials $V(x)$ {\em with parameters inside ${\cal
D}(\alpha,\ell)$} may consistently be chosen as complex
\cite{Carl}.

The main motivation of the detailed study of the domains
exemplified by ${\cal D}(\alpha,\ell)$ lies in their obvious
practical relevance. Virtually no problems emerge in the current
textbook scenario where the explicit specification of the physical
domain of parameters ${\cal D}$ is usually trivial. The standard
choice of the space $I\!\!L_2(I\!\!R)$ and of a manifestly
self-adjoint differential Hamiltonian $H=H^\dagger$ (depending on
some $J$ real couplings or other free parameters) usually enables
us to work with the elementary, unrestricted domain ${\cal
D}\,\equiv\, I\!\!R^J$ of these parameters. In contrast, the
necessary determination of $\partial {\cal D}$ (which plays the
role of a certain horizon of the observability and of the
stability of the quantum system) proved fairly difficult even in
the comparatively elementary example $H^{(DDT})(\alpha,\ell)\neq
\left [H^{(DDT)}(\alpha,\ell)\right ]^\dagger $ as studied in
ref.~\cite{DDT}.

In the widespread terminology coined in refs.~\cite{BB,BBJ} and
employed also in the review \cite{Carl}, the set ${\cal
D}(\alpha,\ell)$ of parameters could be called the domain of the
so called unbroken ${\cal PT}-$symmetry of the system in question
The scope of this specification is adapted to the models
(\ref{SE}) -- usually, one stays inside $I\!\!L_2(I\!\!R)$ and
specifies ${\cal P}$ as the parity reversal and ${\cal T}$ as the
time reversal. Then one requires the ${\cal PT}-$symmetry of the
Hamiltonians $H$ which means that ${\cal PT} H = H {\cal PT}$
\cite{BB,BG}. The choice of the parameters inside ${\cal D}$
(i.e., the physical requirement of the measurability and reality
of the spectrum) is then, finally, rephrased as the so called
${\cal PT}-$symmetry of the wave functions (the review \cite{Carl}
can and should be consulted for all these details).

The nontriviality and the non-smooth, spiked shape of the curve
$\partial {\cal D}(\alpha,\ell)$ as found in~\cite{DDT} opens the
question of a generic characterization of the geometry of
$\partial {\cal D}$ in the less specific setting. In what follows,
we intend to review, extend and {\em complete} our results in this
direction as published in the papers \cite{Hendrik} -- \cite{z}.
In these papers we tried to bridge the apparent gap between the
methods aimed at differential Schr\"{o}dinger operators and at
their alternative matrix representations with $H \neq H^\dagger$.
A supplementary though still sufficiently appealing physical
background of our series of studies has been found in the
anharmonic-oscillator problem with the specific and popular $H =
H^{(AHO)}\neq H^\dagger $ where $V(x) = V^{(AHO)}(x)=x^2 + {\rm
i}g x^3$ or where $V(x)$ has been generalized to an arbitrary real
polynomial in the purely imaginary variable ${\rm i} x$.

The results of refs. \cite{Hendrik} -- \cite{z} (with particular
attention paid to the constructions of the domains ${\cal D}$)
will be reviewed here in sections \ref{sec2} and~\ref{sec22}.
Their extension and completion will be described in sections
\ref{sec2n2} and \ref{sec4}. Section \ref{sec6} is a summary.

\section{\label{sec2}Matrix models with small dimensions }

In the review \cite{Geyer} written in the context of the so called
interacting boson models in nuclear physics, Scholtz et al
emphasized that once a {\em non-unitary} Dyson's fermion-to-boson
mapping is used, the calculation of the energies {\em gets
simplified} while the price to be paid still proves reasonable for
matrices. Indeed, in order to restore the Hermiticity of the
Hamiltonians, just the {\em finite-dimensional} physical Hilbert
space ${\cal H}$ had to be reconstructed {\em ad hoc}. This
observation of merits of working with finite-dimensional matrices
also formed a key encouragement of our forthcoming considerations.

\subsection{Variational inspiration: truncated Hamiltonians
\label{sec2asd}}

By assumption, our starting point given by Schr\"{o}dinger
eq.~(\ref{SE}) with $H = H^{(AHO)}$ remains simple and tractable
in the usual Hilbert space $I\!\!L_2(I\!\!R)$. At the same time,
the necessary transition to the physical Hilbert space ${\cal H}$
(where our complex anharmonic-oscillator $H^{(AHO)}$ becomes
self-adjoint) may remain complicated (interested readers can find
all details in the literature cited in \cite{Carl}). In such a
situation, the use of the basis  ({\em plus its subsequent
variational truncation}) has been recommended in all the series of
our studies \cite{Hendrik} -- \cite{z}. In this way we were able
to reduce the differential-operator Hamiltonians $H^{(AHO)}$ in
eq.~(\ref{SE}) to the sequence of their partitioned matrix
approximate forms
 \ben
 \left (
 \begin{array}{c|c}
 A&B\\
 \hline
 C&D
 \ea
 \right )\ \equiv\
 \left (
 \begin{array}{ccc|ccc}
 H_{1,1}& \ldots& H_{1,K}&
 H_{1,K+1}& \ldots& H_{1,K+K'}\\
 \vdots&&\vdots&\vdots&&\vdots\\
 H_{K,1}& \ldots& H_{K,K}&
 H_{K,K+1}& \ldots& H_{K,K+K'}\\
 \hline
 H_{K+1,1}& \ldots& H_{K+1,K}&
 H_{K+1,K+1}& \ldots& H_{K+1,K+K'}\\
 \vdots&&\vdots&\vdots&&\vdots\\
 H_{K+K',1}& \ldots& H_{K+K',K}&
 H_{K+K',K+1}& \ldots& H_{K+K',K+K'}
 \ea
 \right )\,.
 \een
The truncated basis $\{ |n,\pm\rangle \}$ has been numbered by the
excitation quantum numbers $n=0,1,\ldots$ and it has been
partitioned with respect to the parity $\pm$. For our special
anharmonic cubic $H^{(AHO)}$ this implied that $A$ and $D$ were
real and diagonal, with $A_{k,k} =4k+1$ and $D_{k,k} =4k+3$ in
suitable units. In parallel, all the necessary matrix elements of
the purely imaginary matrices $B={\rm i}B'= C^t={\rm i}(C')^t$
(where $^t$ means transposition) have to be evaluated numerically.
This is partially simplified by the observation that we may
restrict our attention to the purely real matrices since
 \ben
 \left (
 \begin{array}{c|c}
 A&B\\
 \hline
 C&D
 \ea
 \right )\,
 \left (
 \begin{array}{c}
 u\\
 \hline
 v
 \ea
 \right )
 \ \equiv\
 \left (
 \begin{array}{c|c}
 A&B'\\
 \hline
 -C'&D
 \ea
 \right )\,
 \left (
 \begin{array}{c}
 u\\
 \hline
 {\rm i}v
 \ea
 \right )\,.
 \een
Unfortunately, our attempts to construct the domains ${\cal D}$ in
closed form were only successful at $N=K+K'= 2$ \cite{Hendrik}, at
$N=K+K'= 3$ \cite{b} and, with certain surmountable difficulties,
at $N=K+K'= 4$~\cite{c}.

\subsection{Tridiagonal chain models
in the strong-coupling regime \label{sec2b} }

The latter results helped us to imagine that the variations of the
vast majority of the (real) matrix elements of $B'=(C')^t$ did not
lead to any really significant changes in the overall structure of
the respective boundaries $\partial {\cal D}^{(N)}$. For this
reason we further re-arranged the basis and reduced the
variability of the submatrices $B'$ and $C'$ by setting many of
their matrix elements equal to zero. As a net result of all these
more or less natural simplifications we arrived, in ref.
\cite{maximal}, at another family of the matrix toy models
exhibiting the general tridiagonal structure
  \be
 H^{(N)}
 =\left [\begin {array}{cccccc}
  -(N-1)&g_1&&&&\\
 -g_1& -(N-3)&g_{2}&&&\\
 &-g_{2}&\ddots&\ddots&&
 \\
 &&\ddots&N-5&g_{2}&
 \\
 &&&-g_{2}&N-3&g_{1}\\
 &&&&-g_{1}&N-1
 \end {array}\right ]\,.
 \label{hamm}
 \ee
At the dimensions $N=2J$ or $N=2J+1$ these models depend just on a
$J-$plet of real couplings $g_1, g_2, \ldots, g_J$. One of the
most important formal merits of these models is that at any
dimension $N$, all the domain ${\cal D}^{(N)}$ lies inside a
finite hypercube \cite{maximal}. The boundary $\p {\cal D}^{(N)}$
itself can be characterized by its strong-coupling maxima which
were obtained in the following closed form,
 \be
 g_n^{(max)}=\pm (N-n)\,n\,,\ \ \ \ \ n = 1, 2, \ldots, J\,.
 \label{spikes}
 \ee
Although the strong-coupling result (\ref{spikes}) looks easy, its
derivation required extensive computer-assisted symbolic
manipulations. Via a nontrivial extrapolation guesswork we
revealed that geometrically, the horizons $\p {\cal D}^{(N)}$ are
(hyper)surfaces with protruded spikes called extreme exceptional
points, EEPs. This intuitive picture has been complemented by the
more quantitative descriptions of $\p {\cal D}^{(N)}$ in
\cite{II,condit}. It was based on the strong-coupling perturbation
ansatz using an auxiliary, formally redundant small parameter $t$,
 \be
 g_n= g_n^{(max)} \,\sqrt{\left(1-\gamma_n(t) \right) }\,,
 \ \ \ \ \ \ \ \ \
 \gamma_n(t) = t+t^2+\ldots+t^{J-1}+G_n t^J\,.
 \label{lobkov}
  \label{optima}
 \ee
This ansatz extrapolates the rigorous $J\leq 2$ fine-tuning rules
as derived in refs.~\cite{Hendrik,maximal} to all $J$.

\subsection{Secular equations
\label{sec2bbc} }

Once we choose $N=2J$ or $N=2J+1$, abbreviate $E^2=s$ and, at all
the odd dimensions $N=2J+1$, ignore the persistent energy level
$E_{J}^{(2J+1)}=0$, we find that all the secular equations $\det
\left ( H^{(N)} - E \right )=0$ have the same polynomial form,
 \be
 s^J-\left (
 \ba
 J\\1
 \ea
 \right )\,s^{J-1}\,P +\left (
 \ba
 J\\2
 \ea
 \right )\,s^{J-2}\,Q -\left (
  \ba
  J\\3
  \ea
  \right )\,s^{J-3}\,R +
 \ldots = 0\,.
 \label{polyform}
 \ee
At all $J$ and $N$, the coefficients $P, Q, R, \ldots$ should be
understood as  {real} polynomial functions of the $J-$plets of
squares $g_k^2$ of our {real} matrix elements. Once all the
energies are assumed real (i.e., equivalently, once all the roots
$s_k$ of eq.~(\ref{polyform}) happen to be non-negative), we
immediately deduce the following relations tractable as necessary
conditions imposed upon our coefficients in (\ref{polyform}),
 \be
 \ba
 \left (
 \ba
 J\\1
 \ea
 \right )\cdot P
  =
 s_1+s_2+\ldots+s_J\geq 0
  \,,\\
 \left (
 \ba
 J\\2
 \ea
 \right )\cdot Q
  =
 s_1s_2+s_1s_3+\ldots+s_1s_J+s_2s_3+s_2s_4+\ldots +s_{J-1}s_J\geq 0
  \,,\\
 \left (
 \ba
 J\\3
 \ea
 \right )\cdot R
  =
 s_1s_2s_3+s_1s_2s_4+\ldots+s_{J-2}s_{J-1}s_J\geq 0
  \,,\\
 \ldots\ .
  \ea
 \label{glyptoform}
 \ee
In the opposite direction, the set of the necessary inequalities
$P \geq 0$, $Q \geq 0$, $\ldots$ is incomplete as it does not
provide the desirable sufficient condition of observability. It
admits complex roots $s$ in general (take a sample secular
polynomial $(s^2+1)(s-2)$ for illustration). This shows that our
problem of the determination of the physical domains ${\cal
D}^{(N)}$ of couplings is mathematically nontrivial even at the
smallest dimensions $N$ and $J$.

For a given prototype Hamiltonian $H^{(N)}$ and under the
constraints (\ref{glyptoform}), the determination of the domain
${\cal D}^{(N)} = {\cal D}\left ( H^{(N)}\right )$ is {\em
equivalent} to the guarantee of the non-negativity of all the $J$
roots $s_k$  of eq.~(\ref{polyform}). Keeping this idea in mind,
the explicit forms of the corresponding sufficient conditions are
to be given here for the first ten smallest matrix dimensions $N=
2, 3, \ldots, 11$.

\section{The domains ${\cal D}^{(2J)}$ and ${\cal D}^{(2J+1)}$:
a brief review of the known results \label{sec22}}

\subsection{Methodical inspiration:
the non-negativity of the root of eq.~(\ref{polyform}) at
$J=1$ \label{sec2ba} \label{sec2a}}

The first nontrivial illustration of the current {\em Hermitian}
Schr\"{o}dinger's bound-state problem is provided by the
two-by-two real-matrix model
 \ben
 H\,|\psi\rangle = E\,|\psi\rangle\,,\ \ \ \ \
  H=H(a,b,d)=
 \left (
 \begin{array}{cc}
 a&b\\
 b&d
 \ea
 \right )=H^\dagger(a,b,d)\,.
 \een
Its three-parametric spectrum is {\em always} real and, therefore,
observable,
 \ben
 E=E_\pm(a,b,d) =
 \frac{1}{2}\left [
 a+d\pm \sqrt{(a-d)^2+4\,b^2}
 \right ]\,.
 \een
For $H=H(a,b,d)=H^\dagger$ the three-dimensional physical domain
${\cal D}(a,b,d)$ of parameters giving real spectra coincides with
{\em all} $I\!\!R^3$.

The parallel ${\cal PT}-$symmetric two-by-two example is very
similar,
 \ben
 H=H'(a,b,d)=
 \left (
 \begin{array}{cc}
 a&b\\
 -b&d
 \ea
 \right )\,, \ \ \ \ E=E'_\pm(a,b,d) =
 \frac{1}{2}\left [
 a+d\pm \sqrt{(a-d)^2-4\,b^2}
 \right ]\,.
 \een
For each individual choice of the parameters $a$, $b$ and $d$, the
reality of the spectrum $E_\pm'(a,b,d)$  of the primed Hamiltonian
$H'(a,b,d)$ is fragile and it must be guaranteed and proved at a
given triplet of parameters. The reality and stability of the
primed system can only be achieved inside a {\em perceivably
smaller} domain ${\cal D}'(a,b,d)$ with the easily specified EP
horizon,
 \ben
 \p {{\cal D}}'(a,b,d)= \left \{\,
 (a,b,d)\in I\!\!R^3 \ba
 \\
 \ea
 \!\!\right |\, \left .\ba
 \\
 \ea\!\!
 (a-d)^2=4\,b^2
 \right \}\,.
 \een
The interior of the non-compact manifold ${\cal D}'(a,b,d)$ is
specified by the single elementary constraint $ b \in (-|a-d|,
|a-d|)$. This may be interpreted as a fact that the variability of
the parameters $a$ and $d$ is entirely redundant for {\em
qualitative} considerations. It makes sense to get rid of them by
the multiplicative re-scaling of all the parameters and by the
subsequent shift of the energy scale leading to the generic choice
of $a=-1$ and $d=1$.

In the context of section \ref{sec2bbc}, the latter reduction
leads us to the linear version $ s-P=0$ of secular
eq.~(\ref{polyform}) at $J=1$ which has the single root $s_0 = P$.
The non-negativity of this root is equivalent to the
non-negativity of the coefficient $P$. In terms of the single
coupling $g_1=a$ available at $J=1$, the necessary and sufficient
criteria of the observability of $H^{(2)}=H^{(2)}(a)$ or
$H^{(3)}=H^{(3)}(a)$ read $P^{(2)}(a)=1-a^2\geq 0$ and
$P^{(3)}(a)=4-2\,a^2\geq 0$, respectively. In a way transferable
to any dimension, the explicit definitions ${\cal D}^{(2)}(a) =
(-1,1)$ and ${\cal D}^{(3)}(a) = (-\sqrt{2},\sqrt{2})$ may be
re-read as definitions of the corresponding EP horizons $\p {\cal
D}^{(2)}(a) = \{-1,1\}$ and $\p {\cal D}^{(3)}(a)
=\{-\sqrt{2},\sqrt{2}\}$.

In the language of phenomenology, one notices an important
complementarity between the parameter-dependence of the toy
spectra $E_\pm(a,b,d)$ and $E'_\pm(a,b,d)$ at $N=2$. In the former
example, {\em all} of the energies $E_\pm(a,b,d)$ remain safely
real. The second, primed model is less easy to deal with. There
exists the whole set of the eligible two-by-two metric operators
$\Theta =\Theta^\dagger > 0$ which define the inner product in the
corresponding two-dimensional toy Hilbert space ${\cal H}'$ (cf.
\cite{PLB2}). Thus, the operator $H'(a,b,d)$ represents an
observable, in ${\cal H}'$, in spite of its manifest
non-Hermiticity in the auxiliary two-dimensional Hilbert space
${\cal H}$ (where the metric is the Dirac's simplest identity
operator). With parameters inside ${\cal D}'(a,b,d)$ the primed
model remains safely compatible with all the postulates of Quantum
Mechanics, therefore.

\subsection{The non-negativity of all the
roots of eq.~(\ref{polyform}) at $J=2$ \label{sec2c} }

At $J=2$ the quadratic version $ s^2-2\,P\,s^{} +Q=0$ of secular
eq.~(\ref{polyform}) has two roots $s_\pm = P\pm \sqrt{P^2-Q}$.
These two roots remain real if and only if $B \equiv P^2-Q \geq
0$. In the subdomain of parameters where $B \geq 0$ they remain
both non-negative if and only if $P\geq 0$ and $Q \geq 0$. We can
summarize that the required sufficient criterion reads
 \be
 P \geq 0\,,\ \ \ \
 P^2 \geq Q \geq 0\,.
 \label{requi}
 \ee
In an alternative approach, {\em without} an explicit reference to
the available formula for $s_\pm$, let us contemplate the
parabolic curve $y(s)=s^2-2\,P\,s$ which remains safely positive,
in the light of our assumption (\ref{glyptoform}), at all the
negative $s<0$. This curve can only intersect the horizontal line
$z(s)=-Q$ at some non-negative points $s\geq 0$.

The proof of non-negativity of all the roots of our secular
equation degenerates to the proof that there exist two real points
of intersection of the $J=2$ parabola $y(s)$ with the horizontal
line $z(s)$ (which lies below zero) at some $s\geq 0$. Towards
this end we consider the minimum of the curve $y(s)$ which lies at
the point $s_0$ such that $y'(s_0)=0$, i.e., at $s_0=P$. This
minimum must lie {\em below} (or, at worst, at) the horizontal
line of $z(s)=-Q\leq 0$. But the minimum value of $y(s_0)$ is
known, $y(P)=-P^2$. Thus, the condition of intersection $y(s_0)
\leq z(s_0)$ gives the formula $P^2 \geq Q$. QED.

It is amusing to notice that once eq.~(\ref{glyptoform}) holds,
the inequality $P^2-Q\geq 0$ is equivalent to the reality of the
roots simply because $P^2-Q\,\equiv\,(s_1-s_2)^2/4$. Even for some
other two-parametric matrices, precisely {this} type of
requirement is responsible for an important part of the EP
boundary $\p {\cal D}$ (cf. refs.~\cite{b,c} for details).

\section{The domains ${\cal D}^{(2J)}$ and ${\cal D}^{(2J+1)}$:
new results \label{sec2n2}}

The determination of the physical horizons $\p {\cal D}^{(N)}$ of
our models $H^{(N)}$ becomes a more or less purely numerical task
at the very large matrix dimensions $N$ \cite{condit}. In an
opposite extreme, as we already noticed, the non-numerical
exceptions have been found at $N=2$ \cite{Hendrik}, at $N=3$
\cite{b} and at the next two dimensions $N=4$ and
$N=5$~\cite{c,maximal}. Now we intend to complement these
observations by showing that the closed-form constructions of the
prototype horizons $\p {\cal D}^{(N)}$ remain feasible up to the
dimension as high as $N=11$.

\subsection{The non-negativity of all the
roots of eq.~(\ref{polyform}) at $J=3$ \label{sec2d} }

Neither at $N=6$ nor at $N=7$ the sufficient condition of
non-negativity of all the energy roots $s$ is provided by the
three necessary rules $P\geq 0$, $Q\geq 0$ and $R \geq 0$ of
eq.~(\ref{glyptoform}). Let us return, therefore, to the second
method used in paragraph \ref{sec2c} and derive another inequality
needed as a guarantee of the reality of the energies. In the first
step one notices that all the three components of the polynomial
 \ben
 y(s)=s^3-3\,P\,s^2+3\,Q\,s=R\,, \ \ \ \ \ \ \ J=3
 \een
remain safely non-positive at $s<0$. Whenever the roots are
guaranteed real, their non-negativity  $s_n\geq 0$ with $n=1,2,3$
is already a consequence of the three constraints
(\ref{glyptoform}). A guarantee of their reality is less trivial
but it still can be deduced from the shape of the function $y(s)$
on the half-axis $s\geq 0$, i.e., from the existence and
properties of a real maximum of $y(s)$ (at $s=s_-$) and of its
subsequent minimum (at $s=s_+$). At both these points the
derivative $y'(s)=3\,s^2-6\,P\,s+3\,Q$ vanishes so that both the
roots $ s_\pm = P \pm \sqrt{P^2-Q}$ of $y'(s)$ must be real and
non-negative. This condition is always satisfied for the {\em
real} roots $s_k$ of $y(s)$ since
 \ben
 B= P^2-Q\ \equiv\ \frac{1}{54}\,
 \left [
 \left (
 s_1+s_2-2\,s_3
 \right )^2+
 \left (
 s_2+s_3-2\,s_1
 \right )^2+
 \left (
 s_3+s_1-2\,s_2
 \right )^2
 \right ]\geq 0\,.
 \een
In the next step, the sufficient condition of the reality of the
roots $s_k$ will be understood as equivalent to the doublet of the
inequalities $y(s_-)\geq R$ and $y(s_+)\leq R$. Here we may insert
$s_\pm^2=2Ps_\pm-Q$ and get the two inequalities which are more
explicit,
 \be
 2(P^2-Q)\,s_-\ \leq \ PQ-R\ \leq \
 2(P^2-Q)\,s_+\,.
 \label{uhrad}
 \ee
They restrict the range of a new symmetric function of the roots,
 \ben
 P\,Q-R\ \equiv\ \frac{1}{9}\,
 \left [s_1s_2
 \left (
 s_1+s_2-2\,s_3
 \right )+s_2s_3
 \left (
 s_2+s_3-2\,s_1
 \right )+s_3s_1
 \left (
 s_3+s_1-2\,s_2
 \right )
 \right ]\,.
 \een
After another insertion of the known $s_\pm$ we arrive at a
particularly compact formula
 \ben
  2(P^2-Q)^{3/2} \geq R-3\,PQ+ 2P^3 \geq
 -2\,(P^2-Q)^{3/2}\,
 \een
or, equivalently,
 \ben
  4\,\left (P^2-Q \right )^{3} \geq
  \left (R-3\,PQ+ 2P^3\right )^2\,.
 \label{uhradu}
 \een
Due to the numerous cancellations the latter relation further
degenerates to the most compact {missing necessary condition}
 \be
  3\,{P}^{2}{Q}^{2}+6\,RPQ
  \geq 4\,{Q}^{3}+{R}^{2}+4\,R{P}^{3}\,.
 \label{uhrada}
  \ee
Our task is completed. In combination with
eqs.~(\ref{glyptoform}), equation (\ref{uhrada}) plays the role of
the guarantee of the reality of the energy spectrum.

\subsection{The non-negativity of all the
roots of eq.~(\ref{polyform}) at $J=4$ \label{sec2e} }

In a search for the non-negative roots of the quartic secular
equation
 \be
 \det\left ( H^{(8,9)}-E\,I\right ) =
 x^4-4\,P\,x^3+6\,Q\,x^2-4\,R\,x+S\,\equiv\,y(x)+S=0
 \,
 \label{jejichkuba}
 \ee
we note that all the four $N-$dependent coefficients $P$, $Q$, $R$
and $S$ evaluate as polynomials in the squares of the four
coupling parameters $g_k$, $k = 1,2,3,4$. Once all these four
expressions are kept non-negative, the curves $y(x)$ and $z(x)=-S$
do not intersect at $x<0$. At $x \geq 0$ they do intersect four
times at $x\geq 0$ (as required), provided only that the three
extremes of $y(x)$ can be found at the three non-negative real
roots $x_{1,2,3}$ of the extremes-determining equation
 \be
 y'(x_{1,2,3})=4\,(x_{1,2,3}^3-3\,P\,x_{1,2,3}^2
 +3\,Q\,x_{1,2,3}-R)=0\,.
 \label{deset}
 \ee
In an ordering $0 \leq x_1\leq x_2\leq x_3$ of these roots we
arrive at the three sufficient conditions
 \be
 y(x_1)\leq -S\,,\ \ \ \ \
 y(x_2)\geq -S\,,\ \ \ \ \
 y(x_3)\leq -S\,
 \label{jeknub}
 \ee
guaranteeing that the parameters lie inside ${\cal D}^{(8)}$ or
${\cal D}^{(9)}$.

All the three quantities $x_k$ satisfy the cubic equation
$y'(x)=0$ so that its premultiplication by $x$ and enables us to
single out the fourth powers of the roots,
 \ben
 x_{1,2,3}^4=3\,P\,x_{1,2,3}^3-3\,Q\,x_{1,2,3}^2+R\,x_{1,2,3}\,.
 \een
In the other words, we can eliminate all the fourth powers of
these roots from $y(x_{1,2,3})$. This simplification reduces all
the three items in eq.~(\ref{jeknub}) to the three polynomial
inequalities of the third degree,
 \ben
  -P\,x_{1,3}^3+3\,Q\,x_{1,3}^2-3\,R\,x_{1,3}+S \leq 0\,,
  \label{ajejichneknuba1}
 \een
 \ben
  -P\,x_{2}^3+3\,Q\,x_{2}^2-3\,R\,x_{2}+S \geq 0\,.
  \label{ajejichneknuba2}
 \een
Repeating the same elimination of the maximal powers once more, we
may insert
 $
 x_{1,2,3}^3=3\,P\,x_{1,2,3}^2-3\,Q\,x_{1,2,3}+R
 $
and arrive at another equivalent triplet of inequalities
 \begin{eqnarray}
  -
  B\,x_{1}^2
 +2\,B^{3/2}\,C
  \,x_{1} \leq B^2\,D\,,
  \label{bcjejichneknuba2}
 \\
  -
  \label{bcdjejichneknuba2}
  B\,x_{2}^2
 +2\,B^{3/2}\,C
  \,x_{2} \geq B^2\,D\,,\\
  -
  B\,x_{3}^2
 +2\,B^{3/2}\,C
  \,x_{3} \leq B^2\,D\,.
  \label{bjejichneknuba2}
 \end{eqnarray}
The old abbreviations $B=P^2-Q$ and $2\,B^{3/2}\,C=PQ-R$ plus a
new one, $3\,B^2\,D=P\,R-S $ enable us to define
$Y_{1,2,3}:=x_{1,2,3,}/\sqrt{B}$. This yields our final triplet of
the very transparent  quadratic-equation conditions
 \begin{eqnarray}
  Y_{1}^2
 -2\,C
  \,Y_{1} +D\geq 0\,,
  \label{xbcjejichneknuba2}
 \\
    \label{xbcdjejichneknuba2}
  Y_{2}^2
 -2\,C
  \,Y_{2}+D \leq 0\,,\\
  Y_{3}^2
 -2\,C
  \,Y_{3}+D \geq 0\,.
  \label{xbjejichneknuba2}
 \end{eqnarray}
The auxiliary roots $Y_\pm = C \pm \sqrt{C^2-D}$ must be real and
non-negative so that we must guarantee that $D\geq 0$ and $C^2\geq
D$. The conclusion is that eqs.~(\ref{xbcjejichneknuba2}) --
(\ref{xbjejichneknuba2}) degenerate to the four elementary
requirements
 \be
 Y_1\leq Y_-\leq Y_2\leq Y_+\leq
 Y_3\,.
 \label{finalistj}
 \ee
Together with the inequalities  $B\geq 0$, $Q\geq 0$ and $-1 \leq
C-\sqrt{1+Q/B}\leq 1$ they form the final and complete algebraic
definition of the domains ${\cal D}^{(8)}$ and ${\cal D}^{(9)}$.

We can summarize that at $J=4$, the feasibility of the
non-numerical construction of the domains ${\cal D}^{(8)}$ and
${\cal D}^{(9)}$ [determined by eqs.~(\ref{glyptoform}) and
(\ref{finalistj})] is based on the non-numerical solvability of
the third-order polynomial equation~(\ref{deset}).

\subsection{The non-negativity of all the
roots of eq.~(\ref{polyform}) at $J=5$ \label{sec2f} }

Let us finally proceed to  $H^{(N)}$ with $N=10$ and/or $N=11$
which leads to the secular equations of the fifth degree,
 \be
 x^5-5\,P\,x^4+10\,Q\,x^3-10\,R\,x^2+5\,S\,x-T\,\equiv\,y(x)-T=0
 \,.
 \label{zzjejichkuba}
 \ee
From our present point of view the problem of the construction of
the respective horizons $\p {\cal D}^{(N)}$  remains solvable
exactly since the derivative $y'(x)$ is still a polynomial of the
mere fourth degree,
 \be
 \frac{1}{5}\,y'(x)=
 x^4-4\,P\,x^3+6\,Q\,x^2-4\,R\,x+S \,.
 \,
 \label{zzjuba}
 \ee
The exact, real and non-negative values $x_{1}\leq x_2\leq x_3\leq
x_4$ of the four roots of $y'(x)$ (which determine the extremes of
the function $y(x)$) may still be considered available in closed
form.

In a way which parallels our preceding considerations we may
assume that the five $N-$dependent non-negative coefficients
$P\geq 0$, $Q\geq 0$, $R\geq 0$, $S\geq 0$ and $T\geq 0$ obey also
all the additional inequalities derived in the preceding sections.
We may then treat our secular problem (\ref{zzjejichkuba}) as a
search for the graphical intersections between the (nonnegative)
constant curve $z(x)=T$ and the graph of the polynomial $y(x)$ of
the fifth degree (which can only be nonnegative at $x\geq 0$).

Inside the domain ${\cal D}^{(N)}$, the quintuplet of the (unknown
but real and nonnegative) physical energy roots $x_a$, $x_b$,
$x_c$, $x_d$ and $x_e$ has to obey the obvious intertwining rule
 \ben
 0\leq x_a\leq x_1\leq x_b
 \leq x_2\leq x_c\leq x_3
 \leq x_d\leq x_4\leq x_e\,.
 \een
The way towards the sufficient condition of the existence of the
real energy spectrum remains the same as above, requiring
 \be
 y(x_1)\geq T\,,\ \ \ \ \
 y(x_2)\leq T\,,\ \ \ \ \
 y(x_3)\geq T\,,\ \ \ \ \
 y(x_4)\leq T\,.
 \label{zzjeknub}
 \ee
The lowering of the degree should reduce eq.~(\ref{zzjeknub}) to
the quadruplet
 \be
 w(Y_1)\leq 0\,,\ \ \ \ \
 w(Y_2)\geq 0\,,\ \ \ \ \
 w(Y_3)\leq 0\,,\ \ \ \ \
 w(Y_4)\geq 0\,
 \label{wwzzjeknub}
 \ee
where the re-scaling $x_{1,2,3,4}=Y_{1,2,3,4}\,\sqrt{B}$ applies
to the arguments of the brand new auxiliary polynomial function of
the third degree in $Y$,
 \ben
 w(Y)=Y^3-3\,C\,Y^2+3\,D\,Y-G\,.
 \een
Besides the same abbreviations as above, we introduced here a new
one, for $PS-T\,\equiv\,4\,B^{5/2}\,G$. The new and specific
problem now arises in connection with the necessity of finding the
three auxiliary and, of course, real and non-negative roots of the
cubic polynomial $w(Y)$. Once we mark them, in the ascending
order,  by the Greek-alphabet subscripts, we should either
postulate our (in principle, explicit) knowledge of their real and
nonnegative values $Y_\alpha\leq Y_\beta\leq Y_\gamma$ or, in
another perspective, we may recollect simply the above-derived
conditions which restrict the range of the three coefficients $C$,
$D$ and $G$ in the cubic polynomial $w(Y)$.

We may immediately conclude that the last feasible specification
of the domains  ${\cal D}^{(10)}$ and ${\cal D}^{(11)}$ will be
given by the following set of the inequalities,
 \be
 Y_1\leq Y_\alpha\leq Y_2\leq Y_\beta\leq
 Y_3\leq Y_\gamma\leq
 Y_4\,.
 \label{wwfinalistj}
 \ee
This is the desired set of the missing algebraic formulae which
complete the sufficient condition of the reality of the spectra.
We can emphasize that at $J=5$, the feasibility of the present
non-numerical constructions of the most complicated
five-dimensional though still algebraic domains ${\cal D}^{(N)}$
is related again to the most complicated though still
non-numerical solvability of the extreme-determining fourth-order
polynomial equation~(\ref{zzjuba}).

The series of the solvable models is, obviously, exhausted. Any
attempted extension of the recipe beyond $N =11$ would suffer from
the necessity of using mere numerical auxiliary functions of
couplings $g_k$ representing the roots of the extreme-determining
higher-order polynomials.

\section{The wedges of the hypersurfaces $\p {\cal D}^{(N)}$
\label{sec4} }


The existence of the algebraic formulae which determine all the
boundaries $\p {\cal D}^{(N)}$ up to $N=11$ opens a way towards a
verification of the strong-coupling perturbation results of
refs.~\cite{maximal} and \cite{II}. Hopefully, some other similar
qualitative or geometric features of the observability horizons
$\p {\cal D}$ assigned to a given ${\cal PT}-$symmetric
Hamiltonian $H$ will be also detected or discovered via the
reduction of the model to a series of its $N$ by $N$
approximations of the prototype form $H^{(N)}$.

In an alternative setting, we can fix the dimension $N$ and search
for the generic features represented by the given model $H^{(N)}$.
Moving beyond the perturbative framework, for example, our ansatz
(\ref{lobkov}) could be then re-interpreted as a {\em precise}
change of the variables in the space of our free dynamical
parameters. The redundant measure of smallness $t$ may be now
fixed arbitrarily. Instead of the couplings $g_k$ one can also
decide to work with the free parameters $\gamma_k$ or even with
$G_k$s. For illustration one may recollect the two-by-two
Hamiltonian
 \be
 H^{(2)}= \left (
 \begin{array}{cc}
 -1&\sqrt{1-\alpha}\\
 -\sqrt{1-\alpha}&1
 \ea
 \right )\,,\ \ \ \ \alpha \in (0,1)\,
 \label{uno}
 \ee
with the two-point spectrum $E_\pm^{(2)} = \pm \sqrt{ \alpha} \,$.
The variability of the new parameter coincides with ${\cal
D}^{(2)} (\alpha)\ \equiv\ (0,1)$ since there are no additional
constraints.

At the higher dimensions $N$, our horizons $\p {\cal D}^{(N)}$ may
be shown to exhibit a generic hedge-hog-like shape as well as
certain reflection symmetries. They allow us to restrict our
attention to the subdomains of $ {\cal D}^{(N)}$ with the positive
$g_k$s, i.e., with the real quantities $\gamma^{(N)}_J=\alpha$,
$\gamma^{(N)}_{J-1}=\beta$, $\ldots$ which should all remain
non-negative and smaller than one, $\gamma^{(N)}_k\in (0,1)$.

\subsection{New forms of approximations
 \label{sec3aaf} }

Let us now return to the description of the structure of the
boundaries of the domains ${\cal D}^{(N)}$ at $N=6$ and $N=7$ by
means of our key inequality (\ref{uhrad}). In this form of the
rigorous guarantee of the reality of the energies at $J=3$ we may
set $P^2=P^2(B,Q)=B+Q$, postulate $B \geq 0$, $Q \geq 0$ and
insert
 \ben
 s_\pm = \sqrt{B+Q} \pm \sqrt{B} \geq 0\,
 \een
in eq.~(\ref{uhrad}). With an abbreviation
$C:=(PQ-R)/(2\,B^{3/2})\geq 0$ this gives the pair of inequalities
$s_-\ \leq \ C\,{\sqrt{B}} \leq \ s_+$ or, equivalently,
 \be
 {\sqrt{1+q} - 1}\ \leq \ {C}\ \leq \
 {\sqrt{1+q} + 1} \,, \ \ \ \ \ \ \
  q = \frac{Q}{B}\,\in \,(0,\infty)\,.
  \label{uhradeb}
 \ee
Once we notice that
$PQ\,\equiv\,Q\,\sqrt{B+Q}=q\,B^{3/2}\sqrt{1+q}$ we may return
from the auxiliary $C$ to the original $R$ and rewrite
eq.~(\ref{uhradeb}) as our final, perceivably simplified
one-parametric constraint imposed upon the allowed range of the
variability of the value of the polynomial $R$,
 \be
 1+\left ( \frac{q}{2}-1\right )
 \sqrt{1+q}
 \geq
 \frac{R}{2\,B^{3/2}}
 \geq
 \left \{
 \begin{array}{ll}
 0,&q\leq 3\,,\\
 \left ( \frac{q}{2}-1\right )
 \sqrt{1+q}\,-1\,,&q>3\,
 \ea
 \right .\,.
 \label{twosi}
 \ee
In such a reparametrization of the physical domains ${\cal
D}^{(6,7)}$ we use $B\geq 0$ and $Q\geq 0$ and employ our final
form of the two-sided inequality (\ref{twosi}) as a definition of
the allowed range of the remaining quantity $R$. The latter
inequality if particularly strong at the smallest ratios $q=Q/B$.
As long as
 \ben
 1+\left ( \frac{q}{2}-1\right )
 \sqrt{1+q}=
{\frac {3}{8}}{q}^{2}-{\frac {1}{8}}{q}^{3}+{\frac
{9}{128}}{q}^{4}-{ \frac {3}{64}}{q}^{5}+{\frac
{35}{1024}}{q}^{6}+{\cal O}\left ({q}^{7}\right )\,,
 \een
we see that the smallness of $q$ implies the {\em second-order}
smallness of $R$. This means that in the regime where $Q \ll B$,
the three-dimensional physical domains ${\cal D}^{(6,7)}$ are very
narrow in their third dimension represented by $R$.

\subsection{The pairwise
confluences of the levels at $J=3$\label{par3.2} }

For the sake of definiteness let us choose just $N=6$ and
abbreviate
 \ben
 g_1=c\leq \sqrt{5}\,,\ \ \ g_2=b\leq 2\,\sqrt{2}\,,
 \ \ \ g_3=a\leq 3
 \een
in the six-by-six version of matrix (\ref{hamm}). It is easy to
deduce that the domain  ${\cal D}^{(6)}$ is circumscribed by the
ellipsoidal surface given by the equation
 \ben
 P=-\left ({a}^{2}+2\,{b}^{2}+2\,{c}^{2}-35\right )/3=0\,.
 \een
The other two obvious constraints read
 \ben
 3\,Q=
 {b}^{4}+2\,{c}^{2}{a}^{2}-44\,{b}^{2}
 +28\,{c}^{2}-34\,{a}^{2}+{c}^{4}+
 259+2\,{b}^{2}{c}^{2}\geq 0
 \een
and
 \ben
 -R=
 {a}^{2}{c}^{4}-10\,{b}^{2}{c}^{2}
 +30\,{c}^{2}{a}^{2}+225\,{a}^{2}-30\,{c}^{2}
 -{c}^{4}-25\,{b}^{4}-225-150\,{b}^{2}\geq 0\,.
 \een
The last constraint needed to define ${\cal D}^{(6)}$ is then
given by eq.~(\ref{uhrad}).

For illustrative purposes, the latter, purely algebraic
description of the geometric shape of the boundary $\p {\cal
D}^{(6)}$ may be  complemented by the explicit evaluation of all
the six energy levels (say, $E_0\leq E_1 \leq \ldots \leq E_5$).
The easiest answer is obtained at the EEP singular couplings where
the complete confluence of all the energies takes place,  $E_0=
E_1 =\ldots = E_5$.

A clear insight in the structure of the spectrum remains available
in the small EEP vicinity as well. For example, the innermost pair
of the energies $E_2$ and $E_3$ can coincide at $\p {\cal
D}^{(6)}$ (and, subsequently, complexify out of  ${\cal D}^{(6)}$)
while the remaining two doublets remain real, or {\it vice versa}.
In the former scenario we encounter the  pairwise coincidence of
$E_2=E_3=0$ at $\p {\cal D}^{(6)}$. In the latter case we may
abbreviate $E_0=E_1=- 4z=-E_4=-E_5$ and compute the unknown
quantity $z\in (0,1)$.

Alternatively, the two outermost levels [viz.,
$E_0=-E_5=-\sqrt{5\,y}$ where $y \in (0,1)$] can stay real while
the confluence only involves the two internal energy doublets at
the shared values of $E_3=E_4=2\,\sqrt{x}=-E_1=-E_2$ where $x \in
(0,1)$. This is the most complicated example where one has the
relation
 \ben
 (s-s_{max})\,\left [s-s_{min} \right ]^2=
 s^3-
 \left (32\,x^2+25\,y^2 \right )\,s^2
 +\left (
 256\,x^4+800\,x^2y^2
 \right )\,s
 -6400\,x^4y^2=0\,
 \een
which defines a sub-surface of $\p {\cal D}^{(6)}$.

\subsection{The pairwise confluences of the exceptional points
\label{par3.3} }

Let us now return to the $J=3$ option where the surface of the
three-dimensional domain ${\cal D}^{(6)}$ can be visualized as
composed, locally, of the two smooth sub-surfaces which intersect
along a certain double exceptional point (DEP) curve. In terms of
the single free parameter $z$,  the DEP secular equation
degenerates to the formula $
 E^2\,\left (E+4z \right )^2\,\left (E-4z \right )^2=0
$ obtainable from eq.~(\ref{polyform}),
 \be
 s\,\left [s-(4z)^2 \right ]^2=
 s^3-2\,(4z)^2s^2+(4z)^4 s=0\,.
 \label{depka}
 \ee
It is fortunate that the necessary analysis can still be performed
non-numerically since equation~(\ref{depka}) is easy to compare
with the true secular equation (\ref{polyform}) with coefficients
given in paragraph \ref{par3.2}.
%
As long as the factorizable coefficient at $s^0$ must vanish, we
get the first DEP constraint
 $$ \left [a{c}^{2}+15\,a
 + \left (15+{c}^{2}+5\,{b}^{2}\right )\right ]\,
  \left [a{c}^{2}+15\,a
 -\left (15+{c}^{2}+5\,{b}^{2}\right )\right ]
 =0 \,$$
so that we may eliminate
 \ben
 a=\pm \frac{15+{c}^{2}+5\,{b}^{2}}{{c}^{2}+15}\,.
 \een
In the quadrant of the $a-b-c$ space with the positive $a$, the
plus sign must be chosen,
 \ben
 a= 1 +\frac{5\,{b}^{2}}{{c}^{2}+15} \,.
 \een
Thus, we have $3 \geq a \geq 1$ in the closed formula for
 \be
 b^2=\frac{1}{5}\,({c}^{2}+15)\,(a -1)\,
 \label{fei}
 \ee
or, alternatively, for
 \ben
 c^2=\frac{5b^2}{a-1}-15\,.
 \een
This result is to be complemented by the other two relations
  \ben
 3\,Q(c,b,a)=32\,z^2\,,\ \ \ \ \ \ \
 R(c,b,a)=128\,z^4\,.
 \een
A straightforward elimination of $z^2$ gives the second DEP
condition
 \ben
  -66\,{a}^{2}-36\,{b}^{2}+4\,{c}^{2}{a}^{2}-189
  +252\,{c}^{2}-4\,{b}^{2}{a}^{2}-{a}^{4}=0
 \een
with the two compact roots
 \ben
 a^2_{\pm}=
 2\,{c}^{2}-33-2\,{b}^{2}\pm
 2\,\sqrt {{c}^{4}+30\,{c}^{2}-2\,{b}^{2}{c}^{2}
 +225+24\,{b}^{2}+{b}^{4}}\,.
 \een
The acceptable one must be non-negative. For $a^2_-$ this would
mean that $2\,{c}^{2}\geq 33+2\,{b}^{2}$ while, at the same time,
$
 63+12\,{b}^{2}\geq 84\,{c}^{2}
$. These two conditions are manifestly incompatible so that we
must accept the upper-sign root $a^2_+$ which is automatically
positive for all the large $2\,{c}^{2}\geq 33+2\,{b}^{2}$. It also
remains positive for all the smaller $c^2$ constrained by the
requirement
 \ben
 84\,{c}^{2} \geq 63+12\,{b}^{2}\,.
 \een
After the insertion of the definition (\ref{fei}) of $b^2$ we
arrive at the final formula
 \be
 84\,{c}^{2} \geq
 63+{\frac {12}{5}}\,\left (a-1\right )\left
 (15+{c}^{2} \right )\,.
 \label{unequal}
 \ee
It defines a manifestly non-empty domain of parameters at which
one encounters the pairwise confluences of the Kato's exceptional
points.

\section{Summary and outlook \label{sec6} }

There exist two reasons why our knowledge of the physical domains
$ {\cal D}$ is relevant. Firstly, their boundaries $\p {\cal D}$
are marking the breakdown of the reality and observability of the
spectra $\{E_n\}$. Secondly, these boundaries also represent a
regime where the matrices $H^{(N)}$ cease to be diagonalizable.
Thus, it is the {\em simultaneous} degeneracy of the energies and
of the wave functions which characterizes the physics near the
boundary $\p {\cal D}$ of the dynamical domain where the quantum
system starts to become unstable.

\subsection{The onset of instabilities along
$\p {\cal D}$ }

In  Landau's textbook~\cite{Landau} on Quantum Mechanics the
emergence of an instability of a quantum system is exemplified by
a particle in a strongly singular attractive potential
$V(\vec{x})=-G^2/|\vec{x}|^2$. At the critical value
$G_{(max)}^2=-1/4$ of its strength one encounters a horizon beyond
which the particle starts falling on the center. {\it Vice versa},
the system remains stable and physical on all the interval ${\cal
D}= (-1/4, \infty)$ of couplings $G$. From the pragmatic point of
view the Landau's example is not too well selected since the
falling particle should release, hypothetically, an infinite
amount of the energy during its fall. A slightly better textbook
example of the loss of the stability is provided, therefore, by
the Dirac's electron which moves in a superstrong Coulomb
potential. In the language of physics, particle-antiparticle pairs
are created in the system beyond a critical charge
($Z_{(max)}=137$ in suitable units~\cite{Greiner}).

What is shared by the above two {\em Hermitian} sample
Hamiltonians is that they are well defined in a certain domain
${\cal D}$ of parameters  while they lose sense and applicability
for parameter(s) beyond certain horizon. On a less intuitive
level, similar situations have been studied by Kato \cite{Kato}.
He considered several finite-matrix toy Hamiltonians $H(\lambda)$.
He paid attention to the {\em unphysical, complex} values of
$\lambda$ and deduced that the related (in general, complex)
spectra $E_n(\lambda)$ change smoothly with the variation of the
parameter $\lambda$ unless one encounters certain critical,
exceptional points $\lambda^{(EP)}$.

In the above context, certain carefully selected {\em
non-Hermitian} examples seem to be able to offer {\em the best}
illustrative examples in the physical stability analysis. In
addition, our low-dimensional non-Hermitian tridiagonal matrices
also path the way to a combination of mathematics and physics. In
particular, in the context of pure mathematics, our present set of
the solvable examples $H^{(N)}$ with $N \leq 11$ enabled us to
{\em construct and study} the {\em real} version of the Kato's
exceptional points $\lambda^{(EP)}\in \p {\cal D}^{(N)} $. In
parallel, the abundance of the $J$ free parameters in our models
looks more suitable for phenomenological purposes. In particular,
we believe that a systematic characterization of a collapse of a
realistic physical system could make use of these specific models
offering an important link to the possible future classification
and, perhaps, typology of quantum catastrophes \cite{condit}.

\subsection{Degeneracies along the horizons $\p {\cal D}$ }

The obvious theoretical appeal of the problem of stability may be
perceived as one of the explanations of the recent growth of
popularity of the ${\cal PT}-$symmetric and, more generally,
$\eta-$pseudo-Hermitian Hamiltonians  with promising relevance in
quantum field theory \cite{Carl,Mielnik} and in quantum physics in
general \cite{proceedings,webpage}. Although the origin of the
latter ideas can be traced back to the very early days of Quantum
Theory \cite{Mielnik}, the feasibility of its separate
implementations have long been treated as a mere mathematical
and/or physical curiosity (cf., e.g., \cite{BG,referee} for
illustration).

One of the serious technical shortcomings of the ${\cal
PT}-$symmetric and other similar models is that their spectra are
real (i.e., observable) in domains ${\cal D}$ with, sometimes,
very complicated and strongly Hamiltonian-dependent shape of their
EP boundaries $\p {\cal D}$. For an uninterrupted development of
their study it may prove very fortunate that an explicit analytic
description of the horizons $\p {\cal D}$ has been shown available
here for all the matrix ${\cal PT}-$symmetric chain models
$H^{(N)}$ with $N \leq 11$.

In this context it is particularly  important that several recent
microwave measurements \cite{Heiss} confirmed the observability of
the abstract Kato's exceptional points $\lambda^{(EP)}$ in
practice.  These experiments re-attracted attention to the
theoretical analyses of the EP horizons, say, in nuclear physics
where many nuclei can, abruptly, lose their stability
\cite{Geyer,Rotter}. The growth of the role of the EPs may be also
detected in the random-matrix ensembles with various
interpretations \cite{Berry2} and in optical systems (where EPs
are called degeneracies \cite{Berry}). In classical
magnetohydrodynamics the Kato's exceptional points may even happen
to lie {\em inside} the domain of acceptable parameters,
separating merely the different dynamical regimes of the so called
$\alpha^2-$dynamos~\cite{Uwe}.

In all these contexts, our present completion of our recent
studies of EPs may find its future role and relevance as a
classification tool offering a deeper geometric understanding of
the structure of the domains ${\cal D}\left (H \right )$.
Basically, our results seem to indicate an efficiency of a
combination of the methods of algebra (e.g., of solvable
equations) and analysis (offering, e.g., the optimal
parametrizations of elementary curves and (hyper)surfaces) with
the computer-assisted symbolic manipulations and with perturbation
expansions. Perhaps, our explicit verification of the
complementarity, compatibility and productivity of these methods
could also lead, in a not too distant future, to the development
of an explicit control of the stability of the systems, mediated
by some purely algebraic tools of control of parameters in
phenomenological quantum Hamiltonians.

\vspace{5mm}

\section*{Acknowledgements}

Work supported by the GA\v{C}R grant Nr. 202/07/1307, by the
M\v{S}MT ``Doppler Institute" project Nr. LC06002 and by the
Institutional Research Plan AV0Z10480505.


\end{document}